\documentclass[sigconf,authorversion]{acmart}

\AtBeginDocument{%
  \providecommand\BibTeX{{%
    \normalfont B\kern-0.5em{\scshape i\kern-0.25em b}\kern-0.8em\TeX}}}

\copyrightyear{2021} 
\acmYear{2021} 
\setcopyright{acmlicensed}\acmConference[CHI '21]{CHI Conference on Human Factors in Computing Systems}{May 8--13, 2021}{Yokohama, Japan}
\acmBooktitle{CHI Conference on Human Factors in Computing Systems (CHI '21), May 8--13, 2021, Yokohama, Japan}
\acmPrice{15.00}
\acmDOI{10.1145/3411764.3445068}
\acmISBN{978-1-4503-8096-6/21/05}

\usepackage{subfigure}
\usepackage{tabularx}
\begin{document}

%%
%% The "title" command has an optional parameter,
%% allowing the author to define a "short title" to be used in page headers.
\title[Humorous Conversational Agent]{Can a Humorous Conversational Agent Enhance Learning Experience and Outcomes?}

%%
%% The "author" command and its associated commands are used to define
%% the authors and their affiliations.
%% Of note is the shared affiliation of the first two authors, and the
%% "authornote" and "authornotemark" commands
%% used to denote shared contribution to the research.
\author{Jessy Ceha, Ken Jen Lee, Elizabeth Nilsen, Joslin Goh, and Edith Law}
%\authornote{Both authors contributed equally to this research.}
%\email{jceha@uwaterloo.ca}
%\orcid{1234-5678-9012}
%\author{Ken Jen Lee}
%\email{kj24lee@uwaterloo.ca}
%\author{Elizabeth Nilsen}
%\email{enilsen@uwaterloo.ca}
%\author{Joslin Goh}
%\email{jtcgoh@uwaterloo.ca}
%\author{Edith Law}
\email{{jceha, kj24lee, enilsen, jtcgoh, edith.law}@uwaterloo.ca}
\affiliation{%
  \institution{University of Waterloo}
  \streetaddress{200 University Avenue West}
  \city{Waterloo}
  \state{Ontario}
  \country{Canada}
  \postcode{N2L 3G1}
}

% \author{Lars Th{\o}rv{\"a}ld}
% \affiliation{%
%   \institution{The Th{\o}rv{\"a}ld Group}
%   \streetaddress{1 Th{\o}rv{\"a}ld Circle}
%   \city{Hekla}
%   \country{Iceland}}
% \email{larst@affiliation.org}

% \author{Valerie B\'eranger}
% \affiliation{%
%   \institution{Inria Paris-Rocquencourt}
%   \city{Rocquencourt}
%   \country{France}
% }

% \author{Aparna Patel}
% \affiliation{%
%  \institution{Rajiv Gandhi University}
%  \streetaddress{Rono-Hills}
%  \city{Doimukh}
%  \state{Arunachal Pradesh}
%  \country{India}}

% \author{Huifen Chan}
% \affiliation{%
%   \institution{Tsinghua University}
%   \streetaddress{30 Shuangqing Rd}
%   \city{Haidian Qu}
%   \state{Beijing Shi}
%   \country{China}}

% \author{Charles Palmer}
% \affiliation{%
%   \institution{Palmer Research Laboratories}
%   \streetaddress{8600 Datapoint Drive}
%   \city{San Antonio}
%   \state{Texas}
%   \postcode{78229}}
% \email{cpalmer@prl.com}

% \author{John Smith}
% \affiliation{\institution{The Th{\o}rv{\"a}ld Group}}
% \email{jsmith@affiliation.org}

% \author{Julius P. Kumquat}
% \affiliation{\institution{The Kumquat Consortium}}
% \email{jpkumquat@consortium.net}

%%
%% By default, the full list of authors will be used in the page
%% headers. Often, this list is too long, and will overlap
%% other information printed in the page headers. This command allows
%% the author to define a more concise list
%% of authors' names for this purpose.
\renewcommand{\shortauthors}{Ceha, et al.}

%%
%% The abstract is a short summary of the work to be presented in the
%% article.
\begin{abstract}
Previous studies have highlighted the benefits of pedagogical conversational agents using socially-oriented conversation with students. In this work, we examine the effects of a conversational agent's use of affiliative and self-defeating humour --- considered conducive to social well-being and enhancing interpersonal relationships --- on learners' perception of the agent and attitudes towards the task. Using a between-subjects protocol, 58 participants taught a conversational agent about rock classification using a learning-by-teaching platform, the Curiosity Notebook. While all agents were curious and enthusiastic, the style of humour was manipulated such that the agent either expressed an affiliative style, a self-defeating style, or no humour. Results demonstrate that affiliative humour can significantly increase motivation and effort, while self-defeating humour, although enhancing effort, negatively impacts enjoyment. Findings further highlight the importance of understanding learner characteristics when using humour. 

\end{abstract}

%%
%% The code below is generated by the tool at http://dl.acm.org/ccs.cfm.
%% Please copy and paste the code instead of the example below.
%%


%%
%% Keywords. The author(s) should pick words that accurately describe
%% the work being presented. Separate the keywords with commas.

%% A "teaser" image appears between the author and affiliation
%% information and the body of the document, and typically spans the
%% page.

% \begin{teaserfigure}
%   \includegraphics[width=\textwidth]{sampleteaser}
%   \caption{Seattle Mariners at Spring Training, 2010.}
%   \Description{Enjoying the baseball game from the third-base
%   seats. Ichiro Suzuki preparing to bat.}
%   \label{fig:teaser}
% \end{teaserfigure}

%%
%% This command processes the author and affiliation and title
%% information and builds the first part of the formatted document.
\maketitle

\section{Introduction}

Conversational agents---entities with some degree of `intelligence' and the capability of natural language discourse---are becoming a prevalent tool for students within computer-mediated learning environments. Thus, it is essential to understand how best to design these agents to enhance learning experience and outcomes, while taking into account individual characteristics and differences. 
Pedagogical agents have been designed for a number of purposes including tutoring (e.g., \cite{Ward2011}), conversation practice for language learning (e.g., \cite{massaro2006embodied}), and promoting skills such as metacognition (e.g., \cite{ramachandran2018thinking}). Pedagogical agents commonly take on the role of a tutor, co-learner, or novice \cite{roselyn2006case}. This paper focuses on agents in the novice role, also called teachable agents---those with the ability to be taught. These agents are based on learning-by-teaching; a widely studied and practiced technique within the education domain. The approach is inspired by the prot\'eg\'e effect, which demonstrates that learning for the sake of teaching others is more beneficial than learning for one's own self \cite{Biswas2016}.

Previous studies have highlighted the usefulness of pedagogical agents that have socially-oriented conversations with students, e.g., reassuring them \cite{hamam2005learning}, initiating small-talk \cite{Gulz2011}, and engaging in mutual self-disclosure \cite{sinha2015exploring}, leading to more positive experiences and promoting learning gains. Little research exists on the effects of humour in task-oriented human-computer interaction and its use by pedagogical agents. However, the benefits of humour in education have long been alluded to as a result of humour's psychological effects on learners. Humour and laughter have been shown to decrease stress and anxiety, and enhance self-esteem and self-motivation, leading to a more positive emotional and social environment in which students can pay better attention \cite{berk1998, glenn2002brain}. %Further, certain humour styles are more conducive to social well-being and to enhancing interpersonal relationships than others \cite{Martin2003}. 
To our knowledge, there exists no systematic investigation into humour use by pedagogical agents, and, in particular, teachable agents. %We therefore set out to investigate whether we could use social discourse moves by a teachable agent, in particular humour styles known to enhance interpersonal relationships, to promote the prot\'eg\'e effect by enhancing feelings of responsibility between student and agent and by reducing anxiety in the student.
We therefore set out to investigate how varying humour styles in a teachable agent affected participants' perceptions of the agent, experience with the task, and teaching behaviour during the interaction. Further, we explored whether individual characteristics of the learner (in terms of their own humour style) interacted with the humour style of the agent.

The key contributions of this work are:
\begin{itemize}
\item results showing significantly different effects of affiliative, self-defeating, and no humour on participants' perception and effort in teaching an agent, and
\item a discussion of the effects of interactions between the agent's humour style and the participant's own humour style.
%\item design suggestions for implementing humour in teachable agents.
\end{itemize}

%Being intrinsically motivated describes doing an activity for its inherent satisfaction as opposed to being extrinsically motivated by some separate outcome \cite{ryan_deci_2000}. Intrinsic motivation is known to promote exploration, which further fosters curiosity, engagement, active learning, and perseverance \cite{}. Therefore, it can be of interest in education settings to stimulate exploration as an intrinsic motivator of learning. 
%However, although intrinsic motivation is known to promote exploration and perseverance on tasks, anxiety can inhibit it. One possible strategy for alleviating anxiety may lie in the use of humour. 

 %Existing results are contradictory when investigated in different types of agents, such as chatbots, embodied virtual agents, and physical robots. We therefore begin our investigation into humour use by a chatbot.

\section{Related Work}

\subsection{Teachable Agents}
Teachable agents have the ability to be taught and are of particular interest in education, as learning-by-teaching can be a more enriching experience than learning by oneself \cite{duran2017}. In human-human tutoring, the tutor learns more when the tutee struggles with the material, most likely because it leads them to use reflection, self-explanation, and reworking of the problem from multiple angles \cite{walker2009integrating}. Unfortunately, tutee errors do not lead to more learning for the tutee \cite{walker2009integrating} and thus teachable agents are able to capitalize on the beneficial effects for students as peer tutors, while avoiding the detrimental ones of being a peer tutee. Further, research has demonstrated the existence of the prot\'eg\'e effect, by which students are found to try harder to learn for their agent than they do for themselves, and that the social nature of the interaction between student and teachable agent contributes to the effect \cite{chase2009teachable}. Thus, agent characteristics that enhance rapport with the teacher may promote the prot\'eg\'e effect. In the present study, we focus on agent humour.

\subsection{Humour}
%Romero et al. define organizational humour as ``amusing communications that produce positive emotions and cognitions in the individual, group, or organization'' \cite{Romero2006}. 
Humour is an integral part of human communication and can be expressed verbally (produced by means of language or text, e.g., jokes, comics) and/or non-verbally (e.g., facial expression, gesture), intended to elicit responses such as laughter and mirth \cite{martin2018psychology}. Humour and laughter have been shown to decrease stress and anxiety, and enhance self-esteem, self-motivation, and psychological well-being \cite{Kuiper2009,Ford2017,Martin2003}. In education contexts, this effect creates a more positive emotional and social environment in which students can pay better attention \cite{berk1998, glenn2002brain}. Research also demonstrates that material taught with humour can lead to better retention and recall \cite{civikly1986humor, bryant1980relationship, ziv1988teaching, garner2006humor} and humour can increase student interest and ability to engage in divergent thinking \cite{dodge1982heuristics,ziv1988teaching,ziv1983influence}. 

In the field of personality research, humour as a part of an individual's character and the role it plays on the way they engage with others, has generated growing interest. %there has been growing interest in the relationship between an individual's personality and their preferred humour style, in addition to the role that an individual's humour has on the way they engage with others. 
Such work has lead to distinctions between individual differences in humour style---behavioural tendencies related to the uses or functions of humour in everyday life. One of the most prominent contributions in the field distinguishes between four humour styles: \textit{affiliative} (use of benign humour to enhance social relationships), \textit{self-defeating} (use of humour to enhance relationships at one's own expense), \textit{self-enhancing} (use of humour to enhance oneself), and \textit{aggressive} (use of humour to enhance oneself at the expense of others) \cite{Martin2003}. Affiliative and self-defeating humour styles are considered conducive to social well-being and enhancing interpersonal relationships, whereas self-enhancing and aggressive humour are considered as detrimental to social relationships (\cite{Martin2003} p.52). 

The style of humour used can impact the way in which a teacher is perceived. Ziv, Gorenstein, and Moris \cite{ziv1986adolescents} found that students responded differently to a human teacher who used four different types of humour during a lecture: when using a combination of self- and other-disparaging (also known as self-defeating and aggressive) humour, the teacher was rated most appealing and original, when using only other-disparaging humour, the teacher was rated most powerful, and when the teacher did not use humour they were evaluated as having the most systematic teaching style. The study also showed that students who possess a sense of humour are most appreciative of a teacher using humour. Tamborini and Zillman \cite{tamborini1981college} found no difference in rated intelligence when a college lecturer used sexual, other or self-disparaging humour. However, use of self-disparaging humour was perceived as influencing `appeal'. %(an interaction between sex of speaker and sex of respondents was obtained). 
Gruner \cite{gruner1982speaker} showed that speakers using self-disparaging humour were perceived as wittier than those not using humour.

\subsection{Humour in Conversational Agents}
Research on humour in agents has also found the style and form to impact perception and interaction in various ways. In robots, four forms of humour (wit and an ice-breaker, corny jokes, subtle humour, and dry humour and self-deprecation) are suggested to enhance sociality of a robot \cite{kahn2014}, and  innocent humour (riddles and punning jokes) was found to improve perception of task enjoyment, robot personality, and speaking style \cite{Niculescu2013}. In interactions with virtual agents, conversational and situation-specific jokes have been found to affect how cooperation is perceived in an agent \cite{kulms2014}, humour is proposed as a means of recovering from error situations while providing a pleasant user experience \cite{niculescu2015strategies}, and affiliative humour has been shown to significantly motivate healthy behaviours \cite{olafsson2020motivating}.

Morkes, Kernal, and Nass \cite{Morkes1999} found that when participants communicated with a chatbot while working together on a task, but thought they were interacting with another human, the use of humour in on-task conversation lead participants to rate their conversation partner as more likable and reported greater cooperation with, and similarity to, their partner. They also made more jokes and responded more socially. On the other hand, when participants knew they were conversing with a computer agent, the use of humour lead to participants being less social, smiling and laughing less, and feeling less similar to their conversation partner. They also spent less time on the task. Conversely, Dybala et al. \cite{dybala2008humor} reported that participants found a chatbot using puns in off-task conversation to be more natural, interesting, and easier to talk to than a non-humorous agent. In a follow-up study, the researchers also found that the humorous agent was rated as more human-like, funny, and likeable than the non-humorous agent \cite{dybala2009humorized}. The researchers describe two major subclasses of conversational agents that use humour during the dialogue: on-task and off-task. They state: ``we believe that the presence of humour is of higher importance in non-task-oriented agents, for their main purpose is to entertain human interlocutors and socialize with them during the conversation''. Thus, it appears that users are sensitive to an agent's humour, but the impact on interactive behaviour varies depending on the nature of the task.

%Although the use of humour by social conversational agents has become an area of interest, 
Little published research exists on using humour with social agents in learning contexts. One study \cite{wang2010learning} had a virtual human tutor use humorous jokes and pictures in an e-learning interface. In a preliminary experiment they found the humour to enhance learner motivation, performance, and ease emotions. Other research on alternative social expressions points to the possible usefulness of humour use in pedagogical agents. For example, Gulz, Haake, and Silvervarg \cite{Gulz2011} propose that positive learning outcomes after interacting with a teachable agent with off-task social conversational abilities may be due to the off-task social conversation providing an opportunity for cognitive rest, increasing engagement, providing memory cues, and promoting trust and rapport-building between the human and virtual agent. Similarly, Ogan et al. \cite{Ogan2012} suggest that ``agents should keep the student immersed in the experience by making teasing or joking face-threatening moves of their own following an incorrect assessment of their ability''.

\subsubsection{Humour in Teachable Agents}
%We find that in prior studies humour is poorly defined, the results are contradictory, and 
Although prior work indicates the value of humour expression by pedagogical agents, to the best of our knowledge, there has been no previous research systematically investigating the use specifically in agents taking on the role of tutee. %There is also contradictory results in previous work on the effects on perception of a humorous agent. 
We therefore conducted an examination of humour in a teachable agent context. We chose to compare the two humour styles considered conducive to social well-being and enhancing interpersonal relationships: affiliative and self-defeating. These two styles were chosen because: (1) %evidence suggests engaging in off-task conversation during learning tasks has a positive influence on learning experiences and outcomes (e.g., \cite{Gulz2011}), 
learning from a teacher can be considered a social, interpersonal collaboration between teacher and learner \cite{vygotsky1980mind},  and
(2) learning-by-teaching is predicted to be effective as a result of the sense of responsibility the tutor feels for their tutee \cite{chase2009teachable} and we hypothesize that %a self-defeating humour style 
enhancement of the human-agent relationship through humour will increase the sense of responsibility. %, and (3) stress and anxiety may be limiting factors for effort, and humour has been shown to decrease stress and anxiety \cite{glenn2002brain}. %, we hypothesize that the humour %styles, especially affiliative, will decrease stress and increase effort. 
In summary, we predict that humour will enhance the relationship between participant and agent, and decrease stress and anxiety, motivating participants to make more of an effort to teach their agents, and result in a positive effect on learning experience and outcomes.

\section{Study Design}
\subsection{Curiosity Notebook}
\begin{figure}
\centering
\includegraphics[scale=0.55]{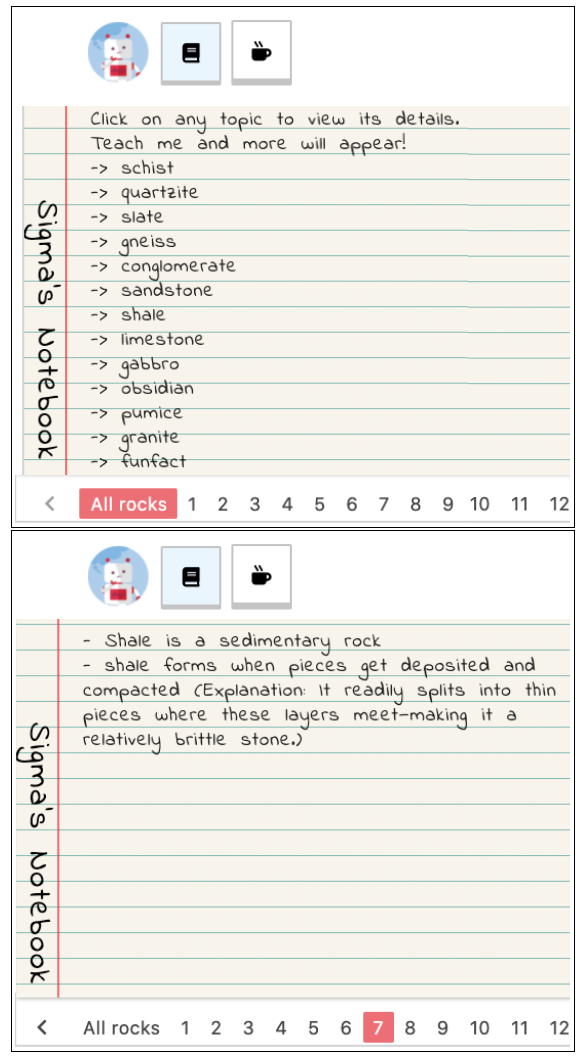}
\caption{Sigma's Notebook, showing list of rocks taught so far (top) and page of notes on Shale rock, with an explanation provided by a participant (bottom).}
\label{fig:notebookpage}
\Description[Sigma's Notebook]{The agent's notebook, with on the top the first page with the list of all rocks that have been taught so far, and on the bottom the page of notes on shale rock, including an explanation given by a participant.}
\end{figure}

\begin{figure*}
\centering
  \includegraphics[scale=0.75]{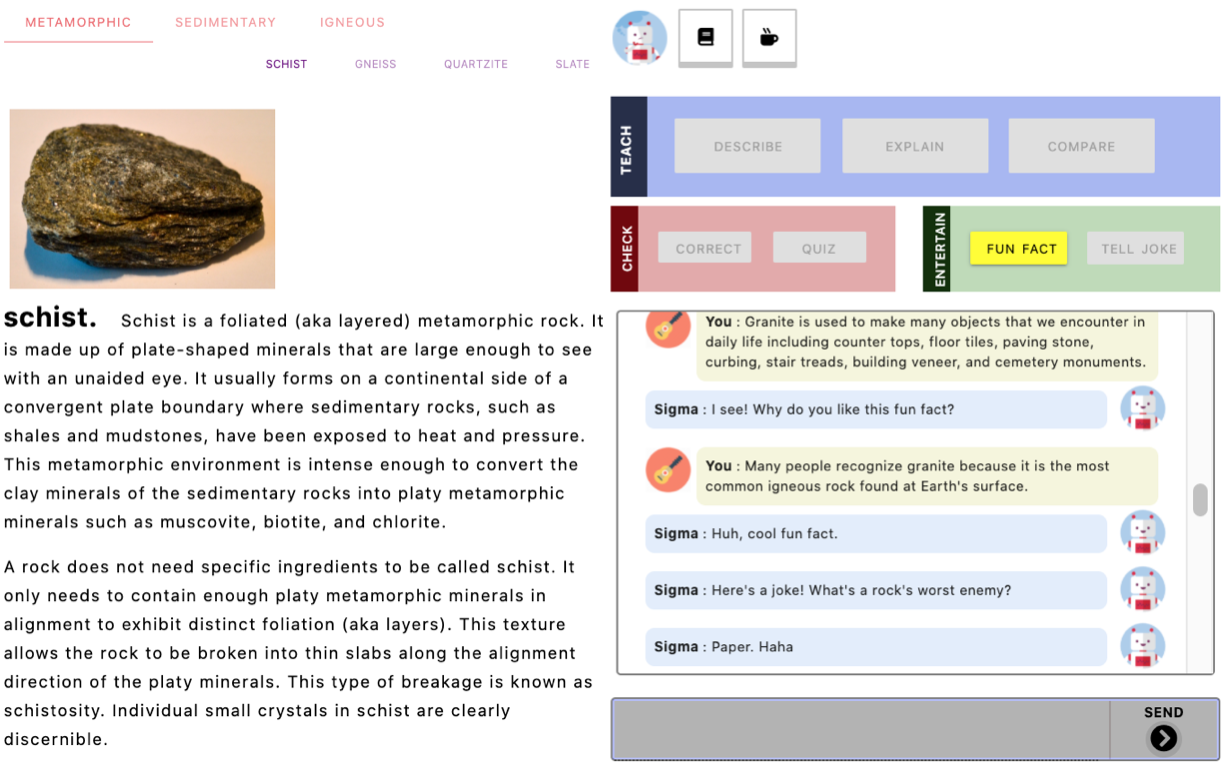}
  \Description[Curiosity Notebook Teaching Interface]{Teaching interface of the Curiosity Notebook application. An article of a schist rock with picture are shown on the left, and the buttons and chat window with an example conversation between participant and agent in the affiliative condition is shown on the right.}
  \caption{Teaching interface, with articles on the left-hand side, chat window on the right, and example interaction (affiliative condition). }~\label{fig:figureTeachingInterface}
\end{figure*}

Using the \textit{Curiosity Notebook} \cite{lawcuriositynotebook2020}, participants taught a virtual agent, Sigma, how to classify rocks. The gender of Sigma was never specified nor implied throughout the entire interaction, and Sigma was represented by a static avatar. 
We adapted and customized the original interface and participant-agent dialogue to fit this study's research questions (For reference, the original layout and task format are described in \cite{lawcuriositynotebook2020}). 
The teaching interface of the Curiosity Notebook (Figure \ref{fig:figureTeachingInterface}) consists of a reading panel on the left-hand side, with a number of articles and pictures about the topic to be taught divided into different categories. Each rock is given its own article. The sentences in the articles could be selected and taught to the agent at certain moments during the interaction. A chat window, through which participants could converse with the agent, was on the right-hand side of the screen. Everything taught to the agent is recorded in a `notebook' that can be viewed at any time (See Figure \ref{fig:notebookpage}). Every rock is given a page in the notebook and the notebook updates live.

% \begin{wrapfigure}{R}{0.65\textwidth}
%   \begin{center}
%     \includegraphics[width=0.63\textwidth]{figures/notebook2.png}
%   \end{center}
%   \caption{Sigma's Notebook, showing page of notes on Shale rock, with an explanation provided by a participant.}~\label{fig:notebookpage}
%   \Description[Sigma's Notebook]{The agent's notebook, with on the left the first page with the list of all rocks that have been taught so far, and on the right the page of notes on shale rock, including an explanation given by a participant.}
% \end{wrapfigure}

The text in the articles is adapted from https://geology.com. The sentences in each article are `linked' to features necessary for classification (large or small crystals, layers, a glassy appearance, holes, sand or pebbles, fossils, and formation process). These linked sentences are used as a ground truth to verify that participants select the correct sentence to teach the agent. If the sentence selected does not match the feature or rock the agent asked about, the agent %expressed  doubt/uncertainty and 
asks the participant to select another sentence that would better answer the question. It also uses this ground truth in answering quiz questions (described in more detail below). %

\subsubsection{Interaction}
When the participant is ready to teach, they select one of seven buttons split into \textit{Teach} (Describe, Explain, or Compare), \textit{Check} (Correct or Quiz) and \textit{Entertain} (Fun Fact or Tell Joke) groupings, which initiates the agent to begin a conversation. Although participants can choose the type of interaction to have with the agent, the agent drives the conversation by asking questions and making statements. 

Describe involves the agent asking for the name of a rock, what rock type it belongs to, and selecting a sentence from the articles with information on what feature(s) or characteristic(s) help classify it. In the Explain interaction, the agent asks why a rock is of a certain type or why it has certain features. Participants answer by selecting a sentence and are sometimes prompted for an explanation in their own words. To select a sentence, the participant is required to navigate to the article of the appropriate rock and find the sentence that contains the answer. Compare has the participant compare two rocks to each other - focusing on what is similar or different depending on whether the two rocks chosen are from the same category or different. 
The Correct button allows participants to change information they had taught the agent. The Quiz button gives participants the opportunity to ask the agent to classify a rock to assess its knowledge state. The Fun Fact button has the agent ask the participant to provide a fun fact and occasionally ask for an explanation of why they thought it was interesting, and the Tell Joke button allows participants to tell the agent a joke. Once the dialogue associated with a button finishes, participants can select another button, allowing them to take a break if needed, and decide what interaction to do next.

\subsection{Experimental Protocol}
We used a between-subjects experimental design, with three conditions: (1) Affiliative, (2) Self-defeating, and (3) Neutral (no verbal humour). Participants were randomly placed into one of the conditions.

\subsubsection{Conditions}
his study focuses on the effect of verbal humour as expressed through humour styles. Although there is a strong connection between humour and laughter, laughter can exist separate from humour. For example, `social/conversational' laughter can be distinguished from `hilarious' laughter, the latter being more directly associated with humour \cite{dupont2016laughter}. Expression of laughter (``haha'') was included in all three conditions for consistency - but not as an expression of verbal humour. %The laughter in all three conditions of our study fill the role of social/conversational laughter.

\textbf{Affiliative} humour is considered relatively benign and self-accepting. Persons with this humour style commonly tell jokes and funny stories for the amusement of others and to facilitate relationships. The agent with an affiliative humour style therefore was designed to occasionally tell jokes throughout the interaction. The jokes were most often conundrum riddles---questions that rely on a play on words in either the question or answer for comedic effect, e.g., ``What's a rock's worst enemy? Paper, haha!'' or ``I've got a joke! What did the one volcano say to the other?... I lava you!''. \textbf{Self-defeating} humour is characterized by an excessive use of self-disparaging humour, by which the user attempts to amuse others at their own expense. The main aim of this style of humour is to achieve social acceptance and approval from others. The agent with a self-defeating humour style interspersed self-disparaging humour throughout the conversation, e.g., ``You know that feeling when you're taught something and understand it right away?... Yeah, not me! Haha!'' or ``When you're a computer but can't learn things by yourself haha''. The agent expressing no verbal humour, \textbf{Neutral}, made statements related to it's self-reflection of learning, e.g., ``This topic is quite interesting!'' or ``Haha I'm enjoying this topic a lot''.

The baseline personality of the agent in all conditions was enthusiastic and curious, saying things like: ``I can always use more information about rocks'', ``Yes please tell me more about rocks!'', ``I want to understand how rocks are formed'', and ``Good idea, let's compare some rocks''. At the end of each interaction associated with a button, %(e.g., Figure ...),
there was an 80\% probability the agent would make an extra statement, reflecting the condition. These statements were always on-topic, i.e., the affiliative jokes were rock-related, the self-defeating jokes were related to the agent's learning about rocks, and the neutral statements were related to the agent's learning about rocks.

\subsubsection{Research Questions}

Based on prior literature from areas of Education, Psychology, Humour, and Conversational Agents, our investigation explores the following research questions:

\begin{itemize}
    \item[\textbf{Q1}:] Does a teachable agent's use of affiliative or self-defeating humour affect participants': 
    \begin{itemize}
    \item perception of the agent, i.e., perceived intelligence, likability, funniness, etc.,
    \item attitudes toward the teaching task, i.e., enjoyment, pressure, motivation, effort etc., and
    \item ability to recall material from the teaching task
\end{itemize}

%\item[\textbf{Q2}:] Do participant characteristics (i.e., humour style, age, gender, education-level, ethnicity, etc.) affect the above measures?
\end{itemize}

According to the Similarity Attraction Hypothesis, people tend to like people they perceive as similar to themselves, and the Media Equation Hypothesis claims that this holds for artificial agents as well \cite{reeves1996media}. Therefore:

\begin{itemize}
    \item[\textbf{Q2}:] Is there an interaction between participant characteristics (i.e., humour style) and the use of humour by a teachable agent, on the above measures?
\end{itemize}

% \textbf{Q3}: Are there certain student characteristics that make the use of humour by a teachable agent more or less effective? (e.g., culture, gender, own sense of humour, own style of humour)

\subsubsection{Humorous Statements}
To generate the humorous statements, six creative writers were recruited through Upwork.  In the task, participants were told to imagine they were a conversational agent that is being taught about classifying rocks. During the conversation they (as the agent) interject different types of humorous statements. For each humour style, they were given the definition, as well as an example interaction between human and agent, and asked to provide 10 statements of each type. Next, a different set of participants (five, also recruited through Upwork) categorized the statements produced by the previous set into the type of humour they felt it belonged. The statements they were shown included those produced by the elicitation Upworkers, as well as `control' statements that did not contain humour, and should therefore be categorized as not belonging to a humour style. The final set of humorous statements were selected by filtering out any that required prior cultural knowledge, were not on the topic of rocks, or were duplicates.

\subsubsection{Procedure} The protocol was conducted entirely online. Participants were sent a link to the Curiosity Notebook, in which they completed all questionnaires as well as interacted with the agent. The system was designed to move participants through each step automatically. Each participant began by reading and signing the information letter and consent form. They then filled in the demographics questionnaire which contained questions on age, gender, cultural background, education, and prior experience with conversational agents. Following this, they completed a pre-study knowledge quiz on rocks. Once submitted, participants were shown a forty second video on how the Curiosity Notebook works and the task participants were expected to complete. They then moved to the teaching interface and were told to set themselves a timer for 40 minutes, after which they should click on the `Stop Teaching' button. Following the interaction, participants completed three questionnaires measuring their perception of the agent, their attitudes and motivation towards the teaching task, and their sense of humour (administered towards the end so as not to prime participants on the focus of humour in the study). The final questionnaire was a post-knowledge quiz to measure their recall of the material. Participants were given a total time of 90 minutes to complete the sequence of surveys and teaching, and at the very end were shown a feedback letter.

\subsubsection{Measures}

We collected data through pre-study and post-study questionnaires, as well as by logging all user interactions on the Curiosity Notebook including, all button and article clicks, conversations between participant and agent, articles discussed, and notes recorded in the notebook.  

To measure participants' attitudes towards the agent we used the Likeability and Perceived Intelligence subscales of the Godspeed questionnaire \cite{bartneck2009}, as well as questions on sense of humour, social ability, and funniness (all presented as semantic differential scales from 1-5), and the Pick-a-Mood pictorial self-report scale for agents \cite{desmet2016mood}. To measure participants' attitudes towards the teaching task we used the Pick-a-Mood pictorial self-report scale for self \cite{desmet2016mood}, and the Interest/Enjoyment and Pressure/Tensions subscales of the Intrinsic Motivation Inventory (IMI) \cite{vallerand_blais_briere_pelletier_1989}. The Academic Motivation Scale (AMS) \cite{vallerand1992} was used to assess the type of motivation elicited by the interaction. The IMI and AMS scales were presented as a Likert scale from 1-7. Pre- and post-knowledge tests were used to measure recall of the material taught. %The Multidimensional Sense of Humour Scale (MSHS), a self-report scale, \cite{thorson1993development} %and 
The Humour Styles Questionnaire (HSQ) \cite{Martin2003}, a self-report scale, was used to measure individual differences in participants' style of humour. %to provide an overall Sense of Humor score. It contains four subscales that distinguish between: (1) humor creativity and uses of humor for social purposes, (2) uses of coping humor, (3) appreciation of humorous people, and (4) appreciation of humor. 
Finally, to measure effort, we used: (1) the Effort/Importance subscale of the IMI \cite{vallerand_blais_briere_pelletier_1989}, and (2) analysis of interaction behaviour while teaching the agent. 

\section{Analysis}

\subsection{Pilot Study}
Eight students took part in a pilot study and received a \$15 Amazon Gift card upon completion. [4 women, 4 men; age range: 20-27 years, mean: 23.1, median: 23]. All participants were undergraduate and graduate students of a research-based university and volunteered for the study by responding to posters. The pilot provided us with some initial results; indicating a clear perception of humour used by the humorous agents versus the non-humorous agent and the ability of participants to move through the procedure seamlessly without researcher involvement. It also made clear the importance of placing the HSQ towards the end of the session to counter-act possible priming effects, as well as having participants complete the post-study knowledge quiz at the very end, so as not to influence their evaluations of the agent based on how well they believe they did on the quiz.

\begin{table*}
\begingroup
\small
\begin{center}
\begin{tabular}{ p{0.2\linewidth} c c c c }
  MAIN STUDY & Neutral(\textit{n}=17) & Affiliative(\textit{n}=18) & Self-Defeating(\textit{n}=18) &  \\
  \hline
 age (years) & \textit{M}=24.8$\pm2.7$ & \textit{M}=25.4$\pm4.4$ & \textit{M}=24.1$\pm3.4$ & \textit{F}(2,50)=0.69, \textit{p} = 0.51  \\  
 \hline
 gender  & 7man & 8man & 5man & $\chi^2 (4, N= 53)$ = 4.68 \\
  & 10woman & 10woman & 11woman & \textit{p} = 0.32 \\
  &  &  & 2non-binary & \\
 \hline
 STEM  & 10yes, 7no & 11yes, 7no & 11yes, 7no & $\chi^2 (2, N= 53)$ = 0.03, \textit{p} = 0.99  \\
 \hline
 native English & 10yes, 7no & 10yes, 8no & 15yes, 3no & $\chi^2 (2, N= 53)$ = 3.68, \textit{p} = 0.16  \\
 \hline
%  Ethnicity & & & & $\chi^2$ = 11.63\\
%  & & & & \textit{p} = 0.66\\
%  Aboriginal or Indigenous & 1 & 0 & 0 &  \\
%  Asian & 7 & 7 & 7 &  \\
%   Black or African American & 0 & 1 & 0 &  \\
%   Hispanic, Latino, or Spanish origin & 1 & 0 & 0 &  \\
%   Middle Eastern or North African & 1  & 1  & 0 &  \\
%   White & 7 & 7 & 9 &  \\
%   White\&Asian & 0 & 1 & 2 &  \\
%   White\&Middle-Eastern & 0 & 1 & 0 &  \\
%  \hline
 Interest in Rocks (1-7)  & \textit{M}=3.18$\pm1.55$ & \textit{M}=3.17$\pm1.69$ & \textit{M}=2.83$\pm1.47$ & $\chi^2 (2, N= 53)$ = 0.59, $p = 0.74$ \\
  \hline
   Knowledge of Rocks (1-7)  & \textit{M}=2.06$\pm0.75$ & \textit{M}=2.39$\pm1.46$ & \textit{M}=2.22$\pm1.44$ & $\chi^2 (2, N= 53)$ = 0.19, \textit{p} = 0.91 \\
 \hline
 Interest in CAs (1-7)  & \textit{M}=4.00$\pm2.00$ & \textit{M}=3.39$\pm1.33$ & \textit{M}=4.50$\pm1.29$ & $\chi^2 (2, N= 53)$ = 4.53, \textit{p} = 0.10 \\
 \hline
 Experience with CAs (1-7) & \textit{M}=3.18$\pm1.94$ & \textit{M}=3.11$\pm1.37$ & \textit{M}=3.50$\pm1.50$ & $\chi^2 (2, N= 53)$ = 0.79, \textit{p} = 0.67 \\
 \hline
%  Affiliative Humour & 43.82 & 46 & 45.94 & p\\
%  Style (max ...) & \pm7.06 & \pm6.25 & \pm6.23 & p \\
%  \hline
%   Self-defeating Humour & 33.59 & 27 & 28.41 & p\\
%  Style (max ...) & \pm9.06 & \pm5.28 & \pm9.01 & p \\
%  \hline
%   Self-enhancing Humour & n & a & sd & p\\
%  Style (max ...) & n & a & sd & p \\
% \hline
%  Aggressive Humour & n & a & sd & p\\
%  Style (max ...) & n & a & sd & p \\

\end{tabular}
\end{center}
\endgroup
\end{table*}

\subsection{Main Study}
\subsubsection{Participants}
58 participants took part in the main study and received a \$15 Amazon Gift card upon completion. [35 women, 21 men, 2 non-binary; age range: 18-35 years, mean: 24.4, median: 25]. All participants volunteered for the study by responding to posters and calls for participation on social media.

\subsubsection{Data Preparation}
Data from 5 participants was removed prior to analysis due to non-compliance with study instructions; therefore results are from \textit{N}=53 (\textit{n}=17 neutral; \textit{n}=18 self-defeating; \textit{n}=18 affiliative). We collected both qualitative and quantitative data from each participant. For the numerous measures the following analyses were carried out: ANOVA, Kruskal-Wallis, linear model, cumulative link model, and stepwise selection method, with condition and demographics (age, humour style, etc.) as the independent factors, and Godspeed, Pick-a-Mood (for self and agent), IMI, AMS, and teaching behaviour as the dependent factors. Interaction effects were investigated, but only those related to participants' humour style (affiliative, aggressive, self-defeating, and self-enhancing; measured by the HSQ \cite{Martin2003}) are reported in this paper.

\section{Results}

% \begin{wrapfigure}{R}{0.45\textwidth}
%   \begin{center}
%     \includegraphics[width=0.43\textwidth]{figures/Post_Mood.png}
%   \end{center}
%   \caption{Results of the Pick-a-Mood pictorial self-report scale for self, across conditions, post-interaction. The number indicates the number of participants that selected each emotion pictogram relating most to themselves.}~\label{fig:selfmood}
%   \Description[Pick-a-Mood for Self]{Graph showing the amount of participants in each condition that picked which mood state. The graph shows the large amount of participants in the self-defeating condition that chose the Neutral mood post-interaction.}
% \end{wrapfigure}

\subsection{Perception of the agents}
\subsubsection{Are the humorous agents more humorous?}
Independent Kruskal-Wallis tests were conducted to examine the differences in the responses to the questions on sense of humour, social ability, and whether the agent was funny or not. No significant differences between the three conditions were found on social ability ($\chi^2 (2, N= 53)$ = 0.16, \textit{p} = 0.93) and funniness ($\chi^2 (2, N= 53)$ = 3.97, \textit{p} = 0.14), but a significant difference ($\chi^2 (2, N = 53)$ = 8.60, \textit{p} = 0.01) was found in rating of sense of humour. Dunn test for multiple comparisons showed both the self-defeating and affiliative conditions differed significantly at $Z= 2.35$, \textit{p} = 0.03 and $Z=2.71 =$, \textit{p} = 0.02, respectively, from the neutral condition. In particular, we noticed that participants (a = affiliative; s = self-defeating; n = neutral) in the humorous conditions perceived the agent to have more of a sense of humour (e.g., ``Sigma had a good sense of humor, which I observed from the jokes they told'' (a16) and ``Humour, expressed through self deprecating jokes'' (s06)), %"Humor through jokes" (a15)) 
than the neutral condition (e.g., ``it would be nice if sigma could tell jokes'' (n09)). %This confirms that our humorous agents were perceived as having a sense of humour, as intended, but interestingly, they were not considered more funny than the non-humorous agent. 

\subsubsection{How do participants feel about the agent's personality?} 
\begin{figure}[b]%
\centering
\subfigure[]{%
\label{fig:robotmood}%
\includegraphics[height=2.5in]{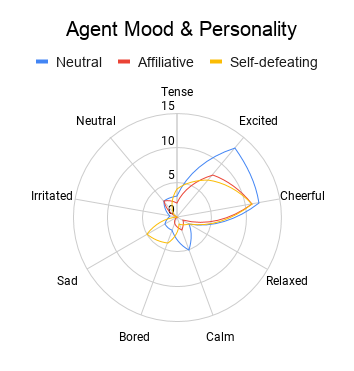}}%
\qquad
\subfigure[]{%
\label{fig:selfmood}%
\includegraphics[height=2.5in]{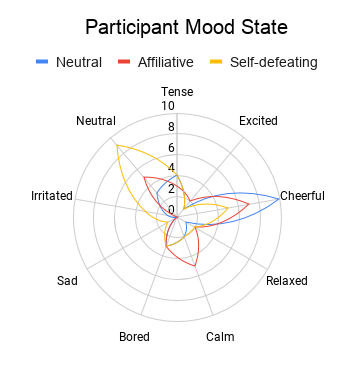}}%
\caption{Results of the Pick-a-Mood pictorial self-report scale for (a) agent's mood and personality, and (b) self, across conditions, post-interaction. The number indicates the number of participants that selected each emotion pictogram.}
\Description[Pick-a-Mood for Self and Agent]{Graphs showing the amount of participants in each condition that picked which state to describe the agent's mood and their own mood. The agent graph shows the sad mood chosen more by participants in the self-defeating condition and the self graph shows the large amount of participants in the self-defeating condition that chose the Neutral mood post-interaction.}
\end{figure}

There were no statistically significant differences between condition means for the Likeability subscale of the Godspeed questionnaire (\textit{F}(2,15) = 0.25, \textit{p} = 0.78). However, there was some evidence that as the participants' self-reported self-enhancing humour style increases, the likeability subscale decreases significantly among those assigned to the self-defeating condition ($\beta = -0.10, t(15)=-1.98, p =0.07$). There was a significant difference found between condition means for the Perceived Intelligence subscale of the Godspeed questionnaire as determined by one-way ANOVA (\textit{F}(2,50) = 3.74, \textit{p} = 0.03). Tukey's honestly significant difference (HSD) post-hoc test showed that both humorous conditions differed significantly at \textit{p} = 0.02; the neutral condition was not significantly different from the humorous conditions. %Figure \ref{fig:perceptionb} shows the average result of the \textbf{pick-a-mood} pictorial self-report scale for the agent. 
The amount the Sad mood pictogram was selected from the Pick-a-Mood pictorial self-report scale to describe the agent's mood and personality was significantly different between conditions ($q(50)=0.29, p=0.02$), i.e., the Sad mood was selected significantly more to describe the agent in the self-defeating condition than in the other two conditions (Figure \ref{fig:robotmood}).

\begin{figure*}
\centering
  \includegraphics[scale=0.55]{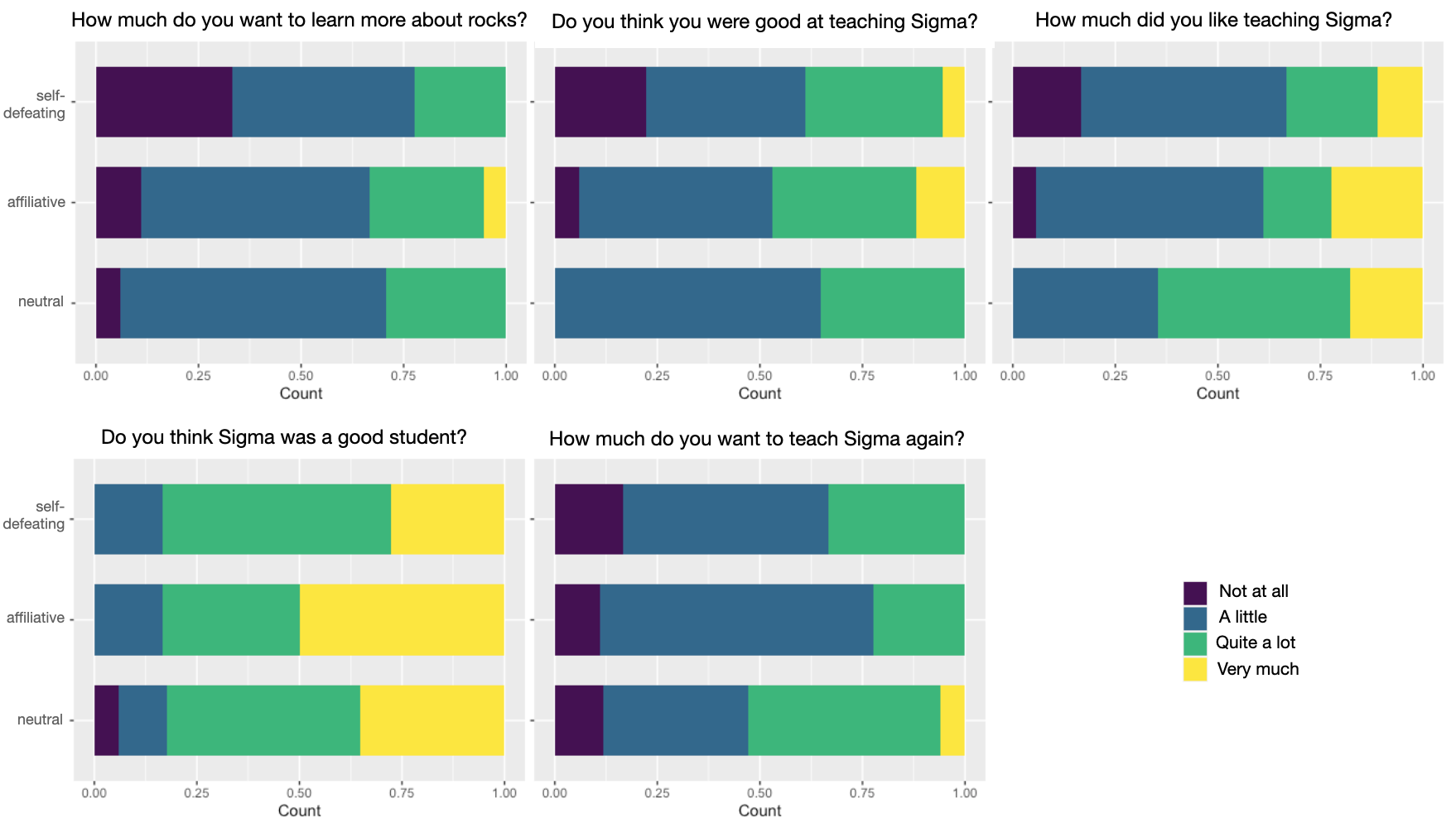}
  \caption{Results of answers to questions on the agent and experience teaching.}~\label{fig:longanswer}
  \Description[Stacked Bar Graphs]{Figure of 5 different 100\% stacked bar graphs showing results of answers ranging from not at all to very much, for each condition on the following questions: How much do you want to learn more about rocks? Do you think you were good at teaching Sigma? How much did you like teaching Sigma? Do you think Sigma was a good student? and How much do you want to teach Sigma again?}
\end{figure*}

Participants were also asked a number of questions relating to experience of the task. The questions and results are shown in Figure \ref{fig:longanswer} and analyzed with cumulative link models (CLM). %Answers to the question: \textit{``How much do you want to teach Sigma again?"} showed no significant differences between conditions. 
The CLM model shows that compared to the other conditions, participants assigned to the self-defeating condition are less likely to have \textit{enjoyed} teaching the agent ($\beta = -1.25, t(48) = -1.92, p = 0.05$). Further, participants in the self-defeating condition with a higher self-enhancing humour style \textit{felt significantly worse} at teaching the agent ($\beta = -0.64, t(21) = -3.05, p = 0.002$), whereas those with a self-defeating humour style themselves in this condition \textit{felt better} at teaching the agent ($\beta = 0.31, t(21) = 1.892, p = 0.06$). %Additionally, participants with a STEM background in the affiliative condition were more likely to want to learn more about rocks after the interaction, $\beta = 8.23, t(19) = 2.53, p = 0.01$. 
Overall, although participants with a higher self-enhancing humour style are significantly \textit{more} likely to think that the agent was a good student ($\beta = 0.23, t(31) = 2.59, p = 0.01$), this probability is \textit{reduced} significantly when the participants are assigned to the affiliative ($\beta = -0.35, t(31) = -2.67, p = 0.01$), or self-defeating condition ($\beta = -0.38, t(31) = -3.09, p = 0.002$). 

\textbf{Teaching Experience} Participants in the self-defeating condition explained: ``Sigma was a bit self-deprecating which wasn't a nice experience being on the teaching side'' (s10), the self-deprecation ``made me feel like I wasn't doing a good job'' %when they kept acting confused/dumb'' 
(s15), and the agent kept making ``jokes about her own incompetence'' (s05). As s15 puts it, the agent had ``a lot of negative talk which was kind of hard to work with. I would have been more encouraged if they were more optimistic about their learning.'' However, self-defeating humour was not always viewed negatively. Four participants in the self-defeating condition explicitly referred to the agent's self-defeating jokes as positive---``I like the jokes Sigma made in between... [the agent] made jokes ... to create a jolly atmosphere'' (s01), ``very positive responses, jokes around and eager to learn, a very easy to teach student'' (s03), ``he tried to lighten the mood with some jokes'' (s13), and ``Sigma was delightful and gave some very human like responses such as self-deprecating jokes that made him feel more like a relative or peer I was teaching rather than a robot'' (s06). Participants also perceived the self-defeating agent as being ``not arrogant at all'' (s06), ``attentive, smart, and friendly'' (s11), ``enthusiastic'' (s12), ``curious'' (s13,s15), and making the experience ``enjoyable and relaxing to read and teach'' (s14).

Participants in the affiliative condition, on the other hand, reflected more positively on the teaching experience in general, with six participants referring to it as being ``fun'' (a01,a06,a07,a13,a17,a18) and ``a cool experience'' (a14). Two participants pointed out their positive opinions of the agent's humour. Participant a04 expressed the agent's ``jokes were a nice added touch :)'' while a16 ``enjoyed the humor''. %On the other hand, a10 found the agent's jokes ``corny'', reflective of humour's subjectivity.

Lastly, participants in the neutral condition mostly referred to the agent as ``friendly'' (n04,n10,n14), ``eager'' (n06,n07,n10,n12,14) and ``curious'' (n03,n09,n18). In terms of teaching, two participants noted the benefit of teaching ``as a good way for me to learn'' (n03) and that they ``get to learn something'' (n01). Three participants reflected positively to the teaching process itself, with n06 stating ``it was fun to try to teach Sigma as much as possible in a short period of time'', n10 explaining ``I enjoyed the task of teaching with the goal of Sigmas success'', and n11 saying ``it was fun trying to decide the best way/which information to teach, plus it was fun learning about rocks in the process.''

\textbf{Comparing Humour}
When asked about the agent's personality traits, five participants in the self-defeating condition listed positive adjectives such as, ``optimistic and positive'' (s08), ``light hearted'' (s06), ``friendly'' (s06,s16,s11), and ``super adorable'' (s13). More participants (nine) in the affiliative condition associated positive attributes, including ``cheerful'' (a01,a14), ``happy'' (a04), ``funny'' (a05,a06,a07,a09), ``a comedian'' (a11), ``brightens your mood'' (a01), and ``pleasant'' (a18). %Coupled with the observations mentioned earlier, it appears that affiliative jokes are generally perceived as being more positive than self-defeating jokes.

% Participants appeared to take more notice of the frequency of jokes told by the agent in the affiliative condition than the self-defeating condition. 
As noted earlier, participants had an 80\% probability of being told a joke by the agent for each button clicked. This probability, however, was perceived as being too frequent by some participants. Six participants in the affiliative condition suggested ``less frequent jokes'' (a12) when asked what they would change about the agent. Telling too many jokes was perceived as ``distracting'' (a17), ``tiresome'' (a15), ``wasted time'' (a08), and the agent being ``not very ... attentive'' (a13). 
On the other hand, for the self-defeating condition, only two participants explicitly referred to the frequency of jokes. Participant s04 felt that the agent should not ``say a joke after every lesson'' as ``it slows things down a little'', and s12 stated the agent made ``too many lame jokes''. Four other participants made more indirect references to the agent's joke frequency. Participant s05 wanted ``less self-deprecation!'', s15 thought the agent should be ``more confident and kind to their self'', and s18 found it ``off-putting'' that ``Sigma made a lot of jokes at its own expense''. Lastly, s10 thought ``it would be nice if Sigma had a more outgoing and nicer personality when interacting''. %and would stray away from self-deprecating humour and maybe instead make more self-motivating comments''.

\subsection{Attitudes towards teaching}
%We used the self-report questions for effort from the IMI and behaviour during the teaching task to measure effort. 

\subsubsection{Did the humorous agents reduce stress/anxiety, and enhance interest and subjective effort?} 
Average scores on the Pressure/Tension, Interest/Enjoyment, and Effort subscales of the IMI questionnaire across conditions had no statistically significant differences between condition means as determined by one-way ANOVA (pressure: \textit{F}(2,50) = 0.83, \textit{p} = 0.44; interest: \textit{F}(2,50) = 0.99, \textit{p} = 0.38; effort: \textit{F}(2,50) = 0.33, \textit{p} = 0.72).  
Figure \ref{fig:selfmood} shows the average result of the Pick-a-Mood pictorial self-report scale for self, across conditions. Post interaction, selection of the Neutral pictogram to describe the participants' own mood state was significantly different between conditions, in particular, participants in the self-defeating condition chose the Neutral mood more than participants in the other conditions.

\subsubsection{Does agent humour use affect user motivation?}
If so, is it intrinsic motivation (IM; actions motivated by the pleasure and satisfaction from the process of engaging in an activity), extrinsic motivation (EM; actions motivated by attaining a goal separate from the process of engaging), or amotivation (AM; the absence of motivation which can co-occur with feelings of low competence)? To investigate this question we used the AMS questionnaire.
% (Figure \ref{fig:ams}). 
The AMS provides overall scores for each type of motivation, and each type is further distinguished into more specific motives: IM - \textit{to know} describes actions performed for the pleasure and satisfaction derived from the learning, exploring, or trying to understand something new from an activity, IM - \textit{toward accomplishment} relates to engaging in actions for the pleasure and satisfaction experienced when trying to achieve something new or beyond one's limits, IM - \textit{to experience stimulation} describes the motivation related to the experiencing of pleasurable sensations, EM - \textit{externally regulated} indicates the behaviour is motivated by reasons external to the task at hand, i.e., payment or rewards, EM - \textit{introjected} refers to actions motivated by pressure an individual puts on themselves, and EM - \textit{identified} describes behaviour that is motivated by the view that participation is important for personal growth \cite{vlachopoulos2005interaction}.

One-way ANOVA showed no significant differences between conditions in the high-level categories of IM $(F(2,50) = 0.84, p = 0.44)$ and AM $(F(2,50) = 0.46, p = 0.64)$, but some weak evidence of a difference in EM $(F(2,50) = 2.52, p = 0.09)$. %, specifically a higher rating in the affiliative condition compared to the self-defeating condition ($p = 0.08$). However, 
There was a significant difference in the average EM - \textit{external regulation} subscale score between conditions $(F(2,50) = 3.92, p = 0.03)$, with Tukey's HSD showing participants in the affiliative condition rating their motivation in the task as externally regulated more highly than participants in the other two conditions (self-defeating-affiliative at $p = 0.03$; and some evidence of significance between affiliative-neutral at $p = 0.08$).
Average AM score was higher for participants with a higher self-reported aggressive humour style in the self-defeating condition ($\beta = 0.18, t(36) = 2.24, p = 0.03$).

% \begin{figure}[b!]
% \centering
%   \includegraphics[scale=0.3]{figures/IM_EM_AM.png}
%   \caption{Comparison of scores on the intrinsic motivation, extrinsic motivation, and amotivation scales of the AMS questionnaire, between conditions. }~\label{fig:ams}
% \end{figure}

\subsubsection{How does agent humour affect effort during teaching?}
As described previously, the Curiosity Notebook allowed participants to interact with the agent in various ways. To measure effort, we recorded the number of total button clicks in the interface, and separated: Teach (Describe, Explain, Compare), Check (Correct, Quiz), and Entertain (Fun Fact, Tell Joke) button clicks, the number of article and category clicks, the frequency of typing out explanations for the agent versus selecting a sentence, as well as the length of these explanations when typed, and amount of time spent teaching. 

Prior work on question-asking has defined numerous schemas for classifying questions based on the effort it would require to find the answers. For example, questions can be classified according to whether the answer can be found in a single, multiple or no sentences in the text \cite{raphael_increasing_1985} or the expected length of an answer and the amount of reasoning required to formulate the answer \cite{graesser1994question}.  
The Curiosity Notebook employs these different types of questions. Within Teach, Describe requires little to no reasoning and the answer can be found in a single sentence in the text, Explain requires slightly more reasoning and the answer can be found in a single sentence (plus participants were sometimes given the opportunity to clarify, in their own words, the selected sentence), and Compare requires more reasoning and the answer is found in multiple sentences and articles. In this way, effort required for the Describe button was less than for the Explain and Compare buttons, which both required more effort. Within Check, Correct involved inspection of the agent's notebook and an understanding of what was incorrect (at times requiring information from multiple sentences), and Quiz involved knowing whether the agent's classification was correct or not (involving only a single sentence). Correct therefore required more effort than Quiz to use. Lastly, of the Entertain buttons, Fun Fact involved selection of a single sentence in the text (at times requiring an explanation of why the fact is fun or interesting) and the Tell Joke button involved no direct interaction with the text. Overall, as the task put to participants was to teach the agent about classifying rocks, we would expect more effort to involve usage of the buttons as follows: Teach > Check > Entertain, and Compare|Explain > Describe > Correct > Fun Fact > Quiz > Tell Joke.

% , and convergent- and divergent-thinking questions,depending on the answers availability in the text \cite{gallagher1963preliminary}.

%Add what each button tells us about effort

Although participants were asked to stop teaching after 40 minutes, it was up to them to finish the interaction after 40 minutes were over. Participants in the affiliative condition decided to spend significantly more time teaching their agents than participants in the other two conditions ($\beta = 8.18, t(49) = 2.58, p = 0.01$). 
The rates at which the buttons were clicked - calculated as number of button clicks divided by the time spent teaching - were analyzed using linear models, where the set of independent variables that can best explain the variance of each measure was selected through stepwise model selection method. %under the Bayesian information criterion (BIC) 
%I focused mostly on descriptive and explanation tasks which I felt I was good at. I don't think I did very well with the comparison tasks. (s18)

\textbf{Teach} There was no significant difference between conditions found in the rate of use of all Teach buttons together ($F(2,50) = 1.46, p = 0.24$), nor the \textbf{Describe} ($F(2,50) = 0.60, p = 0.56$) or \textbf{Explain} ($F(2,50) = 1.21, p = 0.31$) buttons separately. However, participants in the affiliative condition were significantly \textit{less} likely to use the \textbf{Compare} button as frequently as those in the neutral condition ($\beta = -0.007, t(36) =  -3.04, p = 0.004$), but participants in this condition with a higher aggressive or affiliative humour style, were \textit{more} likely to use it ($\beta = 0.0001, t(36) = 2.10, p = 0.04$ and $\beta = 0.0001, t(36) = 2.17, p = 0.04$, respectively). %Participants in the affiliative condition explained their teaching behaviour as: "I explained him the concepts fully" (a05) and "I covered all the concepts of each rock and made sure I explained why each rock was the way they were" (a06).

\textbf{Check} Participants in the self-defeating condition were significantly \textit{less} likely to use the Check buttons (Correct \& Quiz) as frequently as participants in the neutral condition ($\beta = -0.01, t(36) = - 3.03, p = 0.005$). %, e.g., "I was too focused on getting through each sentence of each article and did not finish teaching him about every rock. I also forgot to quiz him" (s14). 
Furthermore, participants with a higher self-reported aggressive humour style were significantly \textit{less} likely to use the Check buttons ($\beta = 0.0002, t(36) = -2.47, p = 0.02$), \textit{unless} they were in the self-defeating condition, and then it was \textit{more} likely ($\beta = 0.0003, t(36) = 3.27, p = 0.002$). Compared to those assigned to the neutral condition, participants with a higher self-reported self-enhancing humour style in the affiliative condition were also \textit{less} likely to use the buttons ($\beta = -0.0002, t(36) = -2.19, p = 0.03$). Looking at the use of the Check buttons separately: %participants across conditions with a higher self-reported affiliative humour style ($\beta = -1.308e04, t(24) = -2.07, p = 0.05$), as well as 
participants assigned to the affiliative condition with a higher self-reported self-enhancing humour style were \textit{less} likely to use the \textbf{Quiz} button as frequently ($\beta = -0.0002, t(24) = -2.38, p = 0.03$). Furthermore, %across conditions, participants with a self-defeating humour style used the \textbf{Correct} button at a higher rate ($\beta = 4.518e-05, t(15) = 2.18, p = 0.05$). P
participants with a higher self-reported aggressive humour style in the affiliative condition were \textit{more} likely to use the \textbf{Correct} button more frequently ($\beta = 0.0001, t(15) = 2.45, p = 0.03$), whereas participants in the same condition with a higher self-reported self-defeating humour style were significantly \textit{less} likely to use it ($\beta = -0.0001, t(15) = -4.23, p = 0.001$). 

\textbf{Entertain} Across conditions, participants with a higher self-reported affiliative humour style were \textit{more} likely to use the buttons ($\beta = 0.0001, t(36) = 2.12, p = 0.04$) unless they were in the self-defeating condition where the probability was \textit{reduced} ($\beta = -0.0002, t(36) = -2.01, p = 0.05$). In contrast, participants with a higher self-reported self-enhancing humour style were \textit{less} likely to use the Entertain buttons ($\beta = -0.00008, t(36) = -2.03, p = 0.05$) unless they were in the self-defeating condition where the probability \textit{increased} ($\beta = 0.0001, t(36) = 2.02, p = 0.05$). Lastly, participants with a higher self-reported aggressive humour style were also \textit{less} likely to use the button ($\beta = -0.0001, t(36) = -2.49, p = 0.02$), across conditions, unless they were in the affiliative condition ($\beta = 0.0001, t(36) = 2.07, p = 0.05$).
Looking at the Entertain buttons separately, the \textbf{Tell Joke} button was not used significantly differently between conditions ($F(2,50) = 1.52, p = 0.23$), nor was the \textbf{Fun Fact} button ($F(2,50) = 0.85, p = 0.44$). 

Participants in the affiliative condition with a higher self-reported affiliative humour style themselves, were \textit{more} likely to check the agent's notebook ($\beta = 0.0004, t(24) = 2.10, p = 0.05$), whereas those in the same condition with a higher self-reported self-enhancing humour style were \textit{less} likely to check it ($\beta = -0.0004, t(24) = -2.34, p =  0.03$). %Participants checked the notebook in order to, for example: ``make sure [the agent was] learning definitive characteristics of each rock type'' (s16). 
The rate of article clicks was significantly \textit{lower} for participants with a higher self-reported affiliative humour style in the self-defeating condition ($\beta = -0.003, t(15) = -2.26, p = 0.04$). Although article clicks can be viewed as a measure of effort, since it can be considered ``too much reading'' (n01), a lower rate may in fact indicate effort as well: ``I read the articles carefully to pick the best response'' (s05). 

There was no significant difference between conditions in the average number of words in explanations written by participants (\textit{F}(2,40) = 0.06, \textit{p} = 0.95), however, there was a significant difference between conditions in whether participants chose to write an explanation in their own words or select a sentence from the articles (\textit{F}(2,50) = 3.56, \textit{p} = 0.04). Tukey's HSD indicated a significant difference between the self-defeating condition and the neutral condition at \textit{p} = 0.04, with participants in the self-defeating condition choosing to type out their own explanations significantly more often than selecting a sentence from the articles. %Examples of written explanations include, ``coming from molten rock material (lava that is really hot), it cools very slowly under the earths surface for a very long time'' (n13), ``When sea creatures die their remains are part of what makes limestone'' (s07), and ``Sandstone looks that way because ... it is made up of particles that have been reduced to sand by weathering'' (a08). Confirming this as a measure of effort, participant n10 stated: ``I found it difficult to explain concepts in my own words'' (n10).

\subsection{Learning}
%See Figure \ref{fig:prepostquiz}. 
Although quiz scores increased post-interaction in all conditions, one-way ANOVA showed no significant difference in change in quiz scores from pre to post interaction ($F(2,50) = 1.12, p = 0.34$), across conditions.

\section{Discussion}

We begin with a discussion of the effects of the humorous agents across all participants, and then take a closer look at how a learner's humour style can interact with that of the agent. 

\subsection{Humorous vs. Non-humorous Teachable Agents}
%In contrast to prior work on humorous conversational agents in off-task interaction (e.g., \cite{dybala2009humorized}), 
\subsubsection{Learning} The humorous agents in our study were not rated more likeable, social, or funny than the non-humorous agent, but they were rated as having more of a sense of humour. %his suggests that participants viewed having a sense of humour as being different from being funny. 
%Martin et al. defined sense of humour as ``the frequency with which the individual smiles, laughs, or otherwise displays amusement in a variety of situations'' \cite{martin1984situational}. As such, having a sense of humour does not necessarily entail that the listener perceives the jokes as funny, it only means the speaker's display of amusement has been acknowledged. 
Humour is successful when both speaker and listener have an obvious intention of amusing each other, whereas failed humour occurs when this intention is unidirectional and the recipient fails to perceive the humour \cite{bogdan2014failed}. This was observed when participants found the agent's jokes to be ``lame'' (s12) or ``corny'' (a10). %``annoying'' (s15). 
In other words, the benefits of having humour in an educational setting might diminish when the humour fails, and could explain the lack of learning gain observed in the study's humorous conditions---humour may only be effective for learning when perceived as funny.

\subsubsection{Experience} Prior work found humour to improve participants' enjoyment of a task \cite{Niculescu2013}, however, in this scenario with a teachable agent, humour did not enhance enjoyment beyond what was experienced by participants in the non-humorous condition, and participants in the self-defeating condition were in fact less likely to have enjoyed teaching their agent. There are several possible reasons why. The first is a self-defeating agent might cause participants to think less of their own competency as the agent's teacher. This is inline with prior research that shows prior student achievement as a valid predictor of collective teacher efficacy \cite{Ross2004} --- the belief that a teacher's efforts can help even the most difficult or unmotivated students \cite{gibson1984teacher}. In other words, the agent's self-defeating jokes could lower participants' own confidence and motivation as a teacher. Importantly, there is a cyclic relationship between student achievement and teacher efficacy \cite{Ross2004};  the agent's self-defeating jokes might cause participants to teach less effectively due to their lowered confidence, for example, participants in the self-defeating condition were less likely to Correct or Quiz their agent than in the neutral condition. Therefore, in the context of learning-by-teaching, self-defeating jokes could negatively impact the benefits. Participants in the self-defeating condition, did however put in more effort in giving the agent their own explanations. This indicates that although effort in the task may increase, it is not accompanied by enjoyment of the learning experience. 

Participants in both humorous conditions noted the overuse of humour. %when the probability of the agent telling a joke after each conversation flow is set at 80\%. 
When asked what, if anything, they would change about the agent, participants in the affiliative condition referred directly to the frequency of jokes (e.g., ``Less frequent jokes!'' (a12)), whereas participants in the self-defeating condition referred more to the agent's personality (e.g., ``More confident and kind to their self'' (s15)). In other words, participants perceived the content of self-defeating jokes as more reflective of the agent's personality than the frequency of affiliative jokes. This is supported by prior research that found affiliative humour to be ``more closely associated with relationship variables than with emotional well-being'', while self-defeating humour is ``related to anxiety, depression, ...and negatively associated with self-esteem and optimism'' \cite{Rnic2016}. This provides a number of insights. First, telling affiliative or self-defeating jokes too frequently may have negatively impacted participants' experience while teaching. Second, the optimal joke frequency is likely different for each participant and dependent on the type of joke, since only six participants in each condition found the agent to be joking too frequently. Third, prior research on associations between humour style and perceived personality applies to teachable conversational agents as well. Similar results on an overuse of humour being perceived as distracting have previously been found \cite{Powell1985,Taylor1974}, and prior work with conversational agents has investigated the timing of jokes, showing that %. Dybala et al. \cite{dybala2010multiagent} developed a multi-agent system which determined when was an appropriate time to tell a pun. The results indicated that 
an agent with appropriately timed humour makes the conversation more interesting than a non-humour-equipped one \cite{dybala2010multiagent}. Future work could look at whether improving timing and amount of affiliative humour style jokes improves the learning experience when conversing with a teachable agent.
 %However, although there was no significant difference between conditions on the IMI Pressure/Tension subscale, which may due to the task not being perceived as stressful, as average rating on this scale across conditions was low, the interaction did result in participants in the affiliative condition describing their mood state as more calm than the other two conditions (although not statistically significant). 

\subsubsection{Motivation} Although the affiliative jokes were sometimes perceived negatively, participants in this condition spent significantly more time teaching their agents - indicating that the negative perception did not impact their willingness to spend time on the task. The humorous agents were generally described as being more human-like than the non-humorous agent, ``almost life like'' (a04), supporting previous work \cite{dybala2009humorized}. Participants in the humorous conditions mentioned the agent ``gave some very human like responses ... more like a relative or peer I was teaching'' (s06) and was ``very personable'' (a12). Meanwhile, participants in the non-humorous condition mostly perceived the agent as a student that needed ``more human like responses'' (n07). Humour made the interaction more engaging and immersive, making it a desirable trait: ``it would be nice if sigma could tell jokes'' (n09). In particular, humour made the agent be perceived less as ``a model student'' (n03) and instead ``give Sigma a personality'' (a03). This human-likeness was hypothesized as a possible contributor to increasing motivation and effort. %however they were less likely to use the Compare button as frequently as those in the neutral condition. We considered the Compare button as one of the buttons which required more effort to use, as it took more time, requiring participants to go to multiple articles and assimilate, compare, or contrast information. However, it may in fact be the case that...
Indeed, our results show that participants in the affiliative condition rated their motivation as externally regulated more highly than participants in the self-defeating condition, %This indicates that behaviour by participants in the affiliative condition was more extrinsically motivated. %and the primary reason individuals are willing to participate in something they are not intrinsically interested in, is because their participation is valued by others to whom they feel connected \cite{ryan_deci_2000}. 
%This provides support for our hypothesis that the use of affiliative humour by the teachable agent aided in the development of a stronger connection between participant and agent, than the non-humorous or self-defeating agents, 
suggesting they were more motivated by the agent than themselves (externally regulated) because of the connection made. Although fostering extrinsic motivation is useful in the short term, as tasks that educators set students are not usually inherently of interest to them \cite{ryan_deci_2000}, the goal of education is commonly to shift behaviour from extrinsically motivated to intrinsically motivated over time \cite{eisenberg1992achievement}. This development over time, as it relates to humour, is worth investigating in the future.

\subsection{Insights for Participants with Different Humour Styles}
As the Similarity Attraction Hypothesis suggests that humans like personalities similar to their own, we expected participants' own humour styles to influence the results. Notable observations regarding participants with certain humour styles are discussed below.

\textbf{Self-Enhancing Humour Style} Research has linked people with a self-enhancing humour style to high self-esteem \cite{Rnic2016} and being more capable of perspective taking empathy \cite{hampes2010relation}. Regardless of condition, participants with more of a self-enhancing humour style rated the agent as being a good student compared to other participants, a sign of higher levels of empathy among them. However, as a self-defeating agent has shown to cause many participants to not enjoy the teaching as much, paired with 
these participants' higher self-esteem, may explain why they were more likely to rate such an agent as a worse student and less likable when compared to other participants and conditions, as well as rating themselves as worse teachers of a self-defeating agent. The higher empathy in participants with this humour style, might further give reason to their observed behaviours during teaching. Across conditions, these participants were less likely to use the Entertain buttons, unless they were in the self-defeating condition in which they were more likely to use them; possibly because of their empathy towards this agent's feelings of not being able to learn well and wanting to ease the agent into the topic. %These characteristics of having a higher self-esteem and more empathy are also reflected in their behaviour in the affiliative condition. 
It is possible that these participants in the affiliative condition were less likely to use the Check buttons, especially quizzing, and checking the agent's notebook, because once they understood how the interface worked, they focused on other tasks. %like %correcting and 
%teaching the agent more fun facts. 
In other words, they needed less feedback and affirmation that the agent was learning what they were teaching, which others might do by repeatedly checking the agent's notebook or quizzing it. 

\textbf{Affiliative Humour Style} Similar to self-enhancing humour, affiliative humour is an adaptive form of humour and linked to increased empathy, however affiliative humour has been found to be more relevant to facilitating relationships and relational functioning %In contrast, self-enhancing humour is more relevant to better emotional well-being among users (higher self-esteem, less anxiety etc.). However, both styles have been linked to increased empathy 
\cite{hampes2010relation}.
% When interacting with a self-defeating agent, participants with more of an affiliative humour style clicked less articles.
When interacting with a self-defeating agent, participants with more of an affiliative humour style were more likely to have a lower article click rate. It is possible that this indicates participants were spending more time reading each article to ensure they taught the agent the most important information in the hopes of improving the agent's perceived sad mood. This may include processes like ``identifying important passages'' (a11) and being able to ``find thesis sentences that would be good for base knowledge'' (a12). This is coupled with the fact that these participants were more likely to use the Entertain buttons across conditions, unless they were teaching the self-defeating agent, where the probability was reduced, i.e., possibly more effort was put into teaching than entertaining. %However, it might also be a sign of lower engagement and therefore participants putting in less effort to teach this type of agent. 
When interacting with the agent that itself had an affiliative humour style, it is possible that the Similarity Attraction Hypothesis encouraged these participants to put in more effort when teaching, observed from the higher probability of checking the agent's notes in the notebook and using the Compare button more frequently (requiring more effort than other button types), than other participants in the affiliative condition.

\textbf{Aggressive Humour Style} As a maladaptive form of humour, the aggressive humour style has been linked to decreased perspective taking empathy and empathetic concern \cite{hampes2010relation}. People with a higher preference of this humour style are also found to be more likely to feel dysphoria and assume others as being more hostile \cite{Rnic2016}. This style is found to be common among students with low school motivation \cite{Saroglou2002}. In other words, participants with more of an aggressive humour style may be more likely to display behaviour reflecting lower motivation in teaching the agent. This is observed in the lower probability of Check type buttons (Quiz; Correct) being used by these participants, which could be explained as a lack of motivation to perform tasks other than teaching the agent. %(e.g., using Teach and Entertain type buttons). 
Their higher amotivation score after interacting with the self-defeating agent also supports the possibility of these participants having lower levels of empathy for the agent. The question then, is how can an agent be designed to increase their motivation? A humorous agent might actually be an answer, as the ability of self-defeating jokes to elicit sympathy might make these participants perceive the agent as less hostile and feel more empathetic towards the agent, and an affiliative humour style may reduce tension during the interaction. % and ultimately are more motivated to teach the agent. 
Indeed, participants with more of an aggressive humour style were significantly more likely to use the Check buttons when they interacted with the self-defeating agent, and when interacting with the agent with an affiliative humour style, these participants %The ability of affiliative jokes to reduce tension during social interactions may also achieve the same effect of increasing these participants' effort, as reflected in the 
had a higher probability of using the Compare and Correct buttons. %being used by participants with a higher aggressive humour style in the affiliative condition compared to those in the neutral condition. 
In other words, both self-defeating and affiliative humour styles in teachable agents showed signs of being able to increase the level of %motivation (and hence effort) 
effort among those with more of an aggressive humour style.

\textbf{Self-Defeating Humour Style} Participants with more of a self-defeating humour style themselves were more likely to rate themselves as being better at teaching when interacting with a self-defeating agent. However, when interacting with an agent with an affiliative humour style they were observed to be less likely to use the Correct button. These observations may be explained by the fact that self-defeating humour is commonly linked to shyness \cite{Hampes2006}, lower self-esteem \cite{Hampes2006,Kuiper2009}, and users are more likely to develop maladaptive social support networks \cite{Kuiper2009}. As such, when interacting with an agent with an affiliative humour style, these participants might feel less capable/worthy of correcting a seemingly confident agent and hence correct it less. In contrast, when interacting with an agent that also displays low self-esteem, they might feel more confident in teaching it, resulting in higher ratings of their own teaching abilities. This explanation might again provide evidence for the Similarity Attraction Hypothesis for the case of the self-defeating humour style, and may also have implications for enhancing student self-efficacy. These insights are of extra importance, since just as the aggressive style, the self-defeating style is typical among students with low school motivation \cite{Saroglou2002}. %As such, these insights can allow for more informed decisions when designing teachable agents for students who are less motivated.

\subsection{Limitations}
Participants in this study were adults between 18-35 years. Although adults in the same age category were used in the humour elicitation stage, this population may differ in terms of humour preference compared to younger children, or older adults, reducing generalizability of the results to these populations. The interaction with the agent was short (one session, of approximately 40 minutes); therefore results could differ for longer exposure. The study focused on only two types of verbal humour, while many more exist, e.g., non-verbal humour or humour styles known as detrimental to interpersonal relationships, as well as being limited to text comprehension and the topic of rock classification. 
% Moreover, more investigation into the interface design is needed. For instance, although the Fun Fact is categorized as an Entertain type button, many participants perceived the Fun Fact button as a way to teach (instead of entertain) the agent.
%if no effect on effort - this could be due to participants' motivation levels being artificially high because it was an experimental setting 
%what happens when a joke fails?
Measuring learning focused on retention rather than a deeper understanding, comprehension, or transfer. Prior work with pedagogical agents has shown that learning with an agent develops a deeper understanding (e.g., \cite{moreno2001case}). Future work could investigate this further using different types of quiz questions to understand what type of learning, is happening. 
Lastly, we acknowledge that humour styles have been found to correlate highly with various other personality characteristics, e.g., empathy, %\cite{hampes2010relation}
self-esteem, optimism, social support, %\cite{Martin2003}, 
and social self-efficacy, %\cite{falanga2014humor}), 
thus we cannot rule out the possibility that we are capturing correlated characteristics (as touched on earlier in the Discussion).
%however our study focuses primarily on participants' own humour style and how it interacted with the agents' humour style (based on the Similarity Attraction Hypothesis), and we found that the similarity attraction effects can be modulated by participants' perception of the agent's behaviour. We touch on some of the correlated characteristics in the Discussion, and call for further research to tease out these factors and their influence on humour's effects. 

\section{Conclusion}
% In prior work, researchers have stated that humour should not be used in on-topic conversation as it can distract students/users \cite{Pawel2009}. They commonly suggest that it should only be used in off-topic dialogue. However, we show that in a teachable agent interaction, on-topic humour may facilitate the learning experience, and that user characteristics can have a large impact on effectiveness.

Although humour has been found to reduce anxiety in students \cite{Martin2003}, increase interest \cite{ziv1983influence}, and create a learning atmosphere in which they can pay more attention \cite{berk1998}, little is known about how it can be used by pedagogical conversational agents. %We hypothesized that enhancing the relationship between participant and agent through humour would motivate participants to make more of an effort in the task, and have a positive effect on their learning experience and outcomes.
Our results indicate that humour, in particular the two humour styles conducive to interpersonal relationships and social well-being, can both enhance, as well as detract from, the experience and outcomes of learners. In general, as an addition to an enthusiastic and curious teachable agent, affiliative humour can increase motivation and effort. Self-defeating humour on the other hand, although increasing effort, does not result in as enjoyable an experience and may cause a decrease in learners' own self-confidence.

We find that humour accentuates the human-likeness of an agent by giving it a personality---e.g., affiliative humour helped demonstrate happiness and intelligence, and self-defeating humour facilitated self-disclosure as the agent displays vulnerability as a struggling learner---possibly resulting in the enhanced effort, motivation, and commitment to the task seen in the humorous conditions. This supports prior work showing that agents that express more human-like qualities, such as relational behavior \cite{bickmore2010}, abilities to build rapport \cite{Ogan2012}, displays of enthusiasm \cite{liew2017exploring}, or sharing vulnerabilities \cite{traeger2020vulnerable,clark2019makes}, can help develop trust and bond between human and agent, making the interaction more engaging and leading to increased motivation. This is especially important in settings where a (teachable) agent is to interact with users over longer periods of time, where having more of a personality can be helpful to increase users' commitment to a task \cite{Morkes1999}.

Some researchers have stated that humour should not be used by agents in on-topic conversation as it can distract users and suggest that it only be used in off-topic dialogue \cite{Pawel2009,dybala2009humorized}, while others have found evidence of the positive influence on learning experiences and outcomes of agents engaging in off-task conversation during learning tasks (e.g., \cite{Gulz2011}). We find that, although on-topic, off-task humour was entertaining and motivating, it was also distracting for some participants and a high-frequency can lead to a loss of enjoyment in a learning-by-teaching scenario. However, the distraction, or extraneous cognitive load, was not necessarily detrimental to affective outcomes, as has been proposed by some researchers \cite{clark2005five}.

Similar to prior work (e.g., \cite{olafsson2020motivating}), our results illuminate the importance of the user's personality characteristics and how they interact with the agent's---indicating that care must be taken in the design of teachable agents, with a one-size-fits-all not always being the most successful when it comes to humour, but can lead to enhanced learning experience and outcomes when matched appropriately. %our results indicate that the effects of humour can be further intensified or dampened by a learners' own humour style.
%This study investigates humour's role in teachable agents and its effects on learner perception of the agent, experience of the task, and teaching behaviour during the interaction.  
Self-defeating jokes appear to evoke empathy and increase effort in learners with more aggressive, self-defeating, or affiliative humour styles, improved the confidence in teaching of learners with a self-defeating humour style themselves, but negatively impacted the experience of learners with more of a self-enhancing humour style. An affiliative humour style similarly increased effort in learners with an aggressive or affiliative humour style, but negatively impacted the effort of learners with a self-enhancing or self-defeating humour style. These findings can be particularly important in informing design decisions for learners with low school motivation, for example, in which maladaptive humour styles are more typical \cite{Saroglou2002}.

Our study highlights the benefits of humour expression by pedagogical conversational agents, especially in the role of a tutee, but also the necessity for more research to investigate the best timing, context, style, and fit so as to make learners relate more to an agent (i.e., via self-defeating humour) and enhance the experience (i.e., via affiliative humour), what effects aggressive and self-enhancing humour styles have, and how combinations of different humour styles may influence learning experiences and outcomes further.

%It is important that you write for the SIGCHI audience. Please read previous years' proceedings to understand the writing style and conventions that successful authors have used. It is particularly important that you state clearly what you have done, not merely what you plan to do, and explain how your work is different from previously published work, i.e., the unique contribution that your work makes to the field. Please consider what the reader will learn from your submission, and how they will find your work useful. If you write with these questions in mind, your work is more likely to be successful, both in being accepted into the conference, and in influencing the work of our field.

\section{Acknowledgments}
%\textit{Anonymized for review}
We thank all participants for their contributions, and acknowledge the funding from NSERC Discovery Grant RGPIN-2015-0454 and University of Waterloo Interdisciplinary Trailblazer Fund for making this work possible.

\bibliographystyle{ACM-Reference-Format}
\bibliography{sample}

% %%
% %% If your work has an appendix, this is the place to put it.
% \appendix

% \section{Research Methods}

% \subsection{Part One}

% Lorem ipsum dolor sit amet, consectetur adipiscing elit. Morbi
% malesuada, quam in pulvinar varius, metus nunc fermentum urna, id
% sollicitudin purus odio sit amet enim. Aliquam ullamcorper eu ipsum
% vel mollis. Curabitur quis dictum nisl. Phasellus vel semper risus, et
% lacinia dolor. Integer ultricies commodo sem nec semper.

% \subsection{Part Two}

% Etiam commodo feugiat nisl pulvinar pellentesque. Etiam auctor sodales
% ligula, non varius nibh pulvinar semper. Suspendisse nec lectus non
% ipsum convallis congue hendrerit vitae sapien. Donec at laoreet
% eros. Vivamus non purus placerat, scelerisque diam eu, cursus
% ante. Etiam aliquam tortor auctor efficitur mattis.

% \section{Online Resources}

% Nam id fermentum dui. Suspendisse sagittis tortor a nulla mollis, in
% pulvinar ex pretium. Sed interdum orci quis metus euismod, et sagittis
% enim maximus. Vestibulum gravida massa ut felis suscipit
% congue. Quisque mattis elit a risus ultrices commodo venenatis eget
% dui. Etiam sagittis eleifend elementum.

% Nam interdum magna at lectus dignissim, ac dignissim lorem
% rhoncus. Maecenas eu arcu ac neque placerat aliquam. Nunc pulvinar
% massa et mattis lacinia.

\end{document}